\documentclass[prb,aps,graphics,floats,showpacs,preprint]{revtex4}

\usepackage{graphicx}
\usepackage{amssymb}
\usepackage{amsmath}
\usepackage{ulem}
\usepackage{color}
\begin{document}

\draft

\title{Tuning magnetotransport in PdPt/Y$_3$Fe$_5$O$_{12}$: Effects of magnetic proximity and spin orbital coupling}
\author{X. Zhou, L. Ma, Z. Shi, and S. M. Zhou$^{*}$\footnotetext{${}^{*}${Correspondence author.
Electronic mail: shiming@tongji.edu.cn}}}
\address{Shanghai Key Laboratory of Special Artificial Microstructure and Pohl Institute of Solid State Physics and School of Physics Science and Engineering, Tongji University, Shanghai 200092, China}

\date{\today}

\vspace{5 cm}

\begin{abstract}

Anisotropic magnetoresistance (AMR) ratio and anomalous Hall conductivity (AHC) in PdPt/Y$_3$Fe$_5$O$_{12}$ (YIG) system are tuned significantly by spin orbital coupling strength $\xi$ through varying the Pt concentration. For both Pt/YIG and Pd/YIG, the maximal AMR ratio is located at temperatures for the maximal susceptibility of paramagnetic Pt and Pd metals. The AHC and ordinary Hall effect both change the sign with temperature for Pt-rich system and vice versa for Pd-rich system. The present results ambiguously evidence the spin polarization of Pt and Pd atoms in contact with YIG layers.
The global curvature near the Fermi surface is suggested to change with the Pt concentration and temperature.
\end{abstract}

\vspace{5 cm}


%----------------------------------------------------------------------------------------------

\maketitle

\indent Generation, manipulation, and detection of pure spin current are popular topic in the community of spintronics because of its prominent advantage of negligible Joule heat in spintronic devices~\cite{D'Yakonov1971,Hirsch1999,Kato2004,Jedema2002,Valenzuela2006}. Pure spin current can be generated by spin Hall effect, spin Seebeck effect (SSE), and etc. By spin Hall effect, the pure spin current can be achieved in semiconductors due to strong spin orbital coupling (SOC). In the SSE approach, the spin current is produced in ferromagnetic materials with a temperature gradient and injected into another nonmagnetic layer through the interface. In general, the pure spin current cannot be probed by conventional electric approach. Instead, it is detected by inverse spin Hall effect~\cite{Ando2008,Saitoh2006}.\\
\indent With strong SOC in Pt layers and long spin diffusion length in Y$_3$Fe$_5$O$_{12}$ (YIG) insulator layers, the Pt/YIG systems are particularly suitable for design and fabrication of spintronic devices~\cite{Czeschka2011,Huang2011,Huang2012,QU2013,Kikkawa2013,Lu2013,Nakayama2013,Vlietstra2013,LuPRB2013,SUN2013}. In studies of the SSE phenomena of Pt/YIG system, the SEE and the anomalous Nernst effect were argued to be entangled~\cite{Huang2011}, where the latter comes from the spin polarization due to the magnetic proximity effect (MPE) of the nearly ferromagnetic Pt layers. Many attempts have been made to study the MPE in Pt/YIG system. Since the atomic magnetic moment of Pt is too small to be measured by magnetometry, anisotropic magnetoresistance (AMR) and anomalous Hall effect (AHE) have been investigated intensively as a function of the Pt layer thickness and sampling temperature ($T$)~\cite{Huang2012,QU2013,Kikkawa2013,Lu2013,Nakayama2013,Vlietstra2013,LuPRB2013}. Up to date, however, magnetotransport results are controversial. The AMR ratio of Pt/YIG system exhibits an angular dependence different from the conventional AMR in magnetic films, and it changes nonmonotonically with the Pt layer thickness. Although these phenomena were attributed to spin Hall magnetoresistance (SMR)~\cite{Nakayama2013,Vlietstra2013} instead of the conventional AMR, the nonmonotonic variation of the AMR with $T$ cannot be understood in the SMR model~\cite{Lu2013}. Moreover, the ferromagnetic ordering in Pt layers was proved by the anomalous Hall effect (AHE) in Pt/YIG system~\cite{Huang2011}. In particular, the mechanism of the observed sign change of the AHE with $T$ is still unclear. Very recently, the x-ray magnetic circular dichroism measurements have been performed by different groups and experimental results are still controversial possibly due to either small atomic magnetic moments of Pt or antiparallel alignment of spins between neighboring Pt atoms~\cite{QU2013,Lu2013,XMCDAPL}. Therefore, an alternative ideal experimental approach must be taken to reveal the MPE in Pt/YIG system. \\
\indent In this work, we will study the SOC effect on the AMR and AHE by using Pd$_{1-x}$Pt$_x$(PdPt)/YIG systems. Here, Pd and Pt atoms are isoelectric elements with different atomic order numbers such that the effective SOC strength can be significantly adjusted by modifying $x$~\cite{Christensen1978}. It is surprising that the AHE and AMR can be tuned significantly for $x$ from 0 to 1.0. Meanwhile, the nonmonotonic variation of the AMR with $T$ is revealed to be caused by the unique $T$ dependence of the spin polarization of Pt and Pd metals. The sign change of AHE with $T$ is found in Pt-rich samples and vice versa in Pd-rich systems. The intriguing phenomena provide strong evidence for the MPE. Meanwhile, the $T$ tuning effects on the curvature near Fermi surface of polarized Pt and Pd layers are illustrated.\\
\indent A series of PdPt/YIG bilayers were fabricated by pulse laser deposition and subsequent magnetron sputtering in ultrahigh vacuum on (111)-oriented, single crystalline Gd$_3$Ga$_5$O$_{12}$ (GGG) substrates. The 70 nm thick YIG thin films were epitaxially grown via pulsed laser deposition from a stoichiometric polycrystalline target using a KrF excimer laser. Secondly, PdPt layers were deposited on YIG thin films by magnetron sputtering. The thickness of the YIG and PdPt layers was determined by the X-ray reflection (XRR) as shown in Fig.~\ref{Fig1}(a). Figure~\ref{Fig1}(b) shows that the x-ray diffraction (XRD) peaks at $2\theta=51$ degrees for (444) orientations in GGG substrate and YIG films overlap each other. The epitaxial growth of the YIG films was confirmed by $\Phi$ and $\Psi$ scan with fixed $2\theta$ for the (008) reflection of GGG substrates and YIG films, as shown in Fig.~\ref{Fig1}(c). In-plane magnetization hysteresis loops of the YIG films were measured at room temperature by vibrating sample magnetometer in Fig.~\ref{Fig1}(d). The measured magnetization of 134 emu/cm$^3$ is almost equal to the theoretical value, and the coercivity is as small as 6.0 Oe. In experiments, half width at half height of the ferromagnetic resonance absorption peak is about 3 Oe~\cite{SUN2013}. Therefore, high quality epitaxial YIG films are achieved in the present work. \\
\indent Before measurements, the films were patterned into normal Hall bar, and then AMR and AHE were measured from 10 to 300 K. Figure~\ref{Fig2}(a) shows the longitudinal resistivity $\rho_{xx}$ versus the external magnetic field $H$ at room temperature. At the saturation state, at the angle between the magnetization and the sensing current $\phi_H=0$ the $\rho_{xx}$ is larger than that of $\rho_{xx}$ at $\phi_H=90$ degrees, similar to the conventional AMR in thick magnetic metallic films such as permalloy~\cite{Lu2013}. Figure~\ref{Fig2}(b) shows the in-plane angular dependence of the AMR at room temperature can be fitted by a linear function of $cos^2\phi_H$, exhibiting a similar attribute in permalloy films. The AMR ratio depends on both the sampling $T$ and $x$, as shown in Fig.~\ref{Fig2}(c). For both Pt/YIG~\cite{Lu2013} and Pd/YIG systems, the $\Delta\rho_{xx}/\rho_{xx}$ shows nonmonotonic variations with $T$; whereas for most ferromagnetic materials it changes monotonically. The maximal value is located near 120 K and 60 K for Pt/YIG and Pd/YIG, respectively. Remarkably,  the susceptibility of paramagnetic Pt and Pd was early observed to have broad peaks almost at the same temperatures~\cite{Gerharadt1981,Inoue1978}. For nonmagnetic transition metals, the enhanced susceptibility $\chi=\chi_0/(1-IN(E_F))$, where $\chi_0$, $I$, and $N(E_F)$ refer to the susceptibility without the presence of Coulomb interaction, the Stoner parameter, and the density of states (DOS) near Fermi level, respectively. Both the Stoner parameter and the MPE induced magnetic moment are also expected to change nonmonotonically~\cite{ZellermannPRL2013,Kublerbook2009}. Therefore, the nonmonotonic dependence of the AMR in Pt/YIG and Pd/YIG systems should stem from their unique $T$ dependence of the induced magnetic moments of both Pt and Pd layers. The monotonic change of the AMR ratio for intermediate $x$ may be due to the contribution of the impurity scattering, as proved below by the large $\rho_{xx}$ at intermediate $x$. It is noted that the present AMR ratio in Pd/YIG is much larger than the values reported by Lin~\textit{et al}~\cite{Lin2013}, possibly due to the weak spin polarization of Pd atoms.\\
\indent In experiments, the Hall resistivity $\rho_{xy}$ was measured as a function of $H$ in the out-of-plane geometry. The anomalous Hall resistivity $\rho_{AH}$ was extrapolated from the linear dependence of $\rho_{xy}$ at large $H$. Figure~\ref{Fig3} shows the Hall loops for Pt/YIG and Pd/YIG at 10 K and 300 K. For Pt/YIG, both $\rho_{AH}$ and the ordinary Hall coefficient $R_0$ are negative at room temperature but positive at 10 K. In contrast, they are negative at both 10 K and 300 K for Pd/YIG samples~\cite{Lin2013}. Figure~\ref{Fig4}(a) shows the $\sigma_{AH}$ as a function of $T$ for all samples, where $\sigma_{AH}=\rho_{AH}/(\rho_{xx}^2+\rho_{AH}*\rho_{xx})\simeq\rho_{AH}/\rho_{xx}^2$ since $\rho_{AH}\ll\rho_{xx}$. The $\sigma_{AH}$ changes from the positive to negative for Pt-rich samples~\cite{Huang2012}; whereas it is always negative in the measured $T$ region for small $x$. Intriguingly, Figure~\ref{Fig4}(b) shows that the $R_0$ also changes the sign for large $x$ whereas no sign change occurs for small $x$. Apparently, the $T$ dependence of $\sigma_{AH}$ is correlated with that of $R_0$ as a function of $x$. Figure~\ref{Fig4}(c) shows at high $T$, $\rho_{xx}$ of all samples increases approximately linearly with $T$ and deviates from the linear dependence at low $T$. The residual resistivity changes nonmonotonically as a function $x$ with a maximum near $x=0.6$, verifying almost random location of Pt and Pd atoms~\cite{Blood1972}.  \\
\indent For spherical Fermi surface, the $R_0$ sign is directly determined by the numbers of electrons and holes. For nonspherical Fermi surface, however, it is also strongly related to the curvature near the Fermi surface. As the integration of the Berry curvature over the Brillouin zone, the intrinsic AHC of magnetic transition metals is naturally determined by the curvature near the Fermi surface. For paramagnetic Pt, the DOS near the Fermi surface changes sharply with the energy~\cite{Kublerbook2009} and the $R_0$ changes the sign~\textit{near} the Fermi level~\cite{Schulz1992}. Due to the exchange splitting and SOC in polarized Pt, not only the numbers of electrons and holes but also the curvature near Fermi surface are significantly different from those of paramagnetic ones~\cite{Campbell1970}. Therefore, the $R_0$ (at low $T$) is positive for polarized Pt (in Fig.~\ref{Fig4}(b)), opposite to that of paramagnetic one~\cite{Dosdale1974,Greig1972,Schulz1992}. With weak exchange splitting and SOC at high $T$, the $R_0$ in polarized Pt is negative, like paramagnetic Pt, as shown in Fig.~\ref{Fig4}(b). With the prominent $T$ effect on the Berry curvature near the Fermi surface, the intrinsic contribution to the $\sigma_{AH}$ in polarized Pt is expected to change the sign with $T$. As well known, the $\sigma_{AH}$ consists of the skew scattering, side-jump, and intrinsic terms~\cite{Nagaosa2010}. Since the magnitude of the skew scattering term (proportional to $\sigma_{xx}$) changes slightly with $T$, the $\sigma_{AH}$ for Pt-rich systems is also expected to change the sign, as shown in Fig.~\ref{Fig4}(a). The \textit{co-occurrent}
sign changes of both $R_0$ and $\sigma_{AH}$ strongly verify the globally varying curvature near the Fermi surface. In contrast, for Fe and Mn$_5$Ge$_3$ films, the $R_0$ changes from the negative to positive near 80 K whereas the $\sigma_{AH}$ is always positive below room temperature, and the sign change is attributed to the change of the conductivity ratio of \textit{d} and \textit{s} bands instead of the global curvature change near the Fermi surface~\cite{Klaffy1974,Cottam1968,Zeng2006}. For NiPt thin films, the $\sigma_{AH}$ rather than $R_0$ changes the sign with $T$, which is attributed to other reasons rather than the global change of the curvature near the Fermi surface~\cite{Golod2013}. Due to weak SOC in polarized Pd, the global curvature near the Fermi surface changes less prominently, compared with that of paramagnetic one and therefore $R_0$ in the measuring $T$ region is always negative, like the paramagnetic one. Meanwhile, neither $R_0$ nor $\sigma_{AH}$ changes the sign with $T$ as shown in Figs.~\ref{Fig4}(a)$\&$~\ref{Fig4}(b). Similarly, for pure Ni films neither $R_0$ nor $\sigma_{AH}$ changes the sign below room temperature~\cite{McAlister1979}. The present correlation between the $\sigma_{AH}$ and $R_0$ with the Pt concentration verifies that they are largely determined by the curvature near the Fermi surface. The $T$ tuning effect on the electronic band structure is also demonstrated in Pt/YIG. \\
\indent Significant SOC effect on the magnetotransport properties in PdPt/YIG is illustrated. Figures~\ref{Fig5}(a)$\&$~\ref{Fig5}(b) show the $\Delta\rho_{xx}/\rho_{xx}$ and $\sigma_{AH}$ at 10 K as a function of $x$, respectively. The AMR ratio is $8\times 10^{-4}$ and $1\times 10^{-4}$ for Pt/YIG and Pd/YIG bilayers, respectively. It is enhanced in magnitude by a factor of about one order from $x=0$ to 1.0. In principle, the AMR in ferromagnetic materials arises from the \textit{s}-\textit{d} scattering, and it is theoretically predicted to be proportional to the square of the SOC strength $\xi^2$ if the resistivity ratio of spin-up and spin-down channels is fixed according to the perturbation theory~\cite{Malozemoff1986}. Since the $\xi$ of Pt is about 3 times that of Pd~\cite{Christensen1978}, it is experimentally proved that the ratio of the AMR between Pt/YIG and Pd/YIG is close to that of $\xi^2$. For intermediate $x$, the AMR ratio deviates from the quadratic dependence due to large contribution of the impurity scattering as shown in Fig.~\ref{Fig5}(c). With increasing $x$, the $\sigma_{AH}$ at 10 K changes from the negative to positive. For Pt/YIG and Pd/YIG, it is about 3.0 and -1.0 (S/cm), respectively, and the magnitude ratio is close to that of $\xi$ between two elements~\cite{Christensen1978,Zeng2006}. Therefore, the sign and magnitude of $\sigma_{AH}$ in PdPt/YIG system are tuned by changing $\xi$ with various $x$. \\
\indent In summary, the AMR ratio in PdPt/YIG system can be enhanced by a factor of about one order from $x=0$ to $x=1$. It changes nonmonotonically with $T$ due to similar $T$ dependence of the atomic magnetic moment. At 10 K, the AHC magnitude of $x=1$ is about 3 times that of $x=0$. For Pt-rich samples both the $R_0$ and AHC change their signs with $T$ and vice versa for Pd-rich system, due to the global change of the curvature near the Fermi surface with $x$. The SOC tuning effects on the magnetotransport properties can be understood based on the perturbation theory. All present phenomena directly evidence the MPE in the PdPt/YIG. The $T$ tuning effect on the electronic band structure is also demonstrated in polarized PdPt layers. The present work will also be helpful for optimizing the spintronics devices. \\
\indent {\bf Acknowledgments} This work was supported by the National Science Foundation of China Grant Nos.11374227, 51331004, 51171129, and 51201114, the State Key Project of Fundamental Research Grant No.2009CB929201, and Shanghai Nanotechnology Program Center (No. 0252nm004).\\
% as well as by the National Science Council of Taiwan and the Center for Quantum Science and Engineering, National Taiwan University (CQSE-10R1004021).\\

\begin{figure}[p]
\begin{center}
FIGURE CAPTIONS
\end{center}

\begin{flushleft}
Figure 1 (color online): For typical Pt/YIG films, x-ray reflectivity at small angles (a) and XRD diffraction at large angles (b), $\Phi$ and $\Psi$ scan with fixed $2\theta$ for the (008) reflection of GGG substrate and YIG film (c). In (d) is shown the room temperature in-plane magnetization hysteresis loop of the YIG layer. In (a) black and red lines correspond to YIG and Pt layers, respectively.
\end{flushleft}

\begin{flushleft}
Figure 2 (color online): For Pt$~$(1 nm)/YIG films, AMR curves at $\phi_H=0$ and 90 degrees (a) and angular dependent AMR at $H=10$ kOe (b). For PdPt$~$(1 nm)/YIG films, the AMR ratio versus $T$ for various $x$ (c).
\end{flushleft}

\begin{flushleft}
Figure 3 (color online): For Pt$~$(1 nm)/YIG (a, b) and Pd$~$(1 nm)/YIG (c, d) films, $\rho_{xy}$ versus $H$ at 10 K (a, c) and 300 K(b, d).
\end{flushleft}

\begin{flushleft}
Figure 4 (color online): For PdPt$~$(1 nm)/YIG films, $\sigma_{AH}$ (a), $R_0$ (b), and $\rho_{xx}$ (c) versus $T$ for various $x$.
\end{flushleft}

\begin{flushleft}
Figure 5 (color online): For PdPt$~$(1 nm)/YIG films, AMR (a), $\sigma_{AH}$(b), and $\rho_{xx}$ (c) at 10 K versus $x$. Solid lines serve a guide to the eye.
\end{flushleft}
\end{figure}

\begin{figure}[p]
\begin{center}
\resizebox*{15 cm}{!}{\includegraphics*{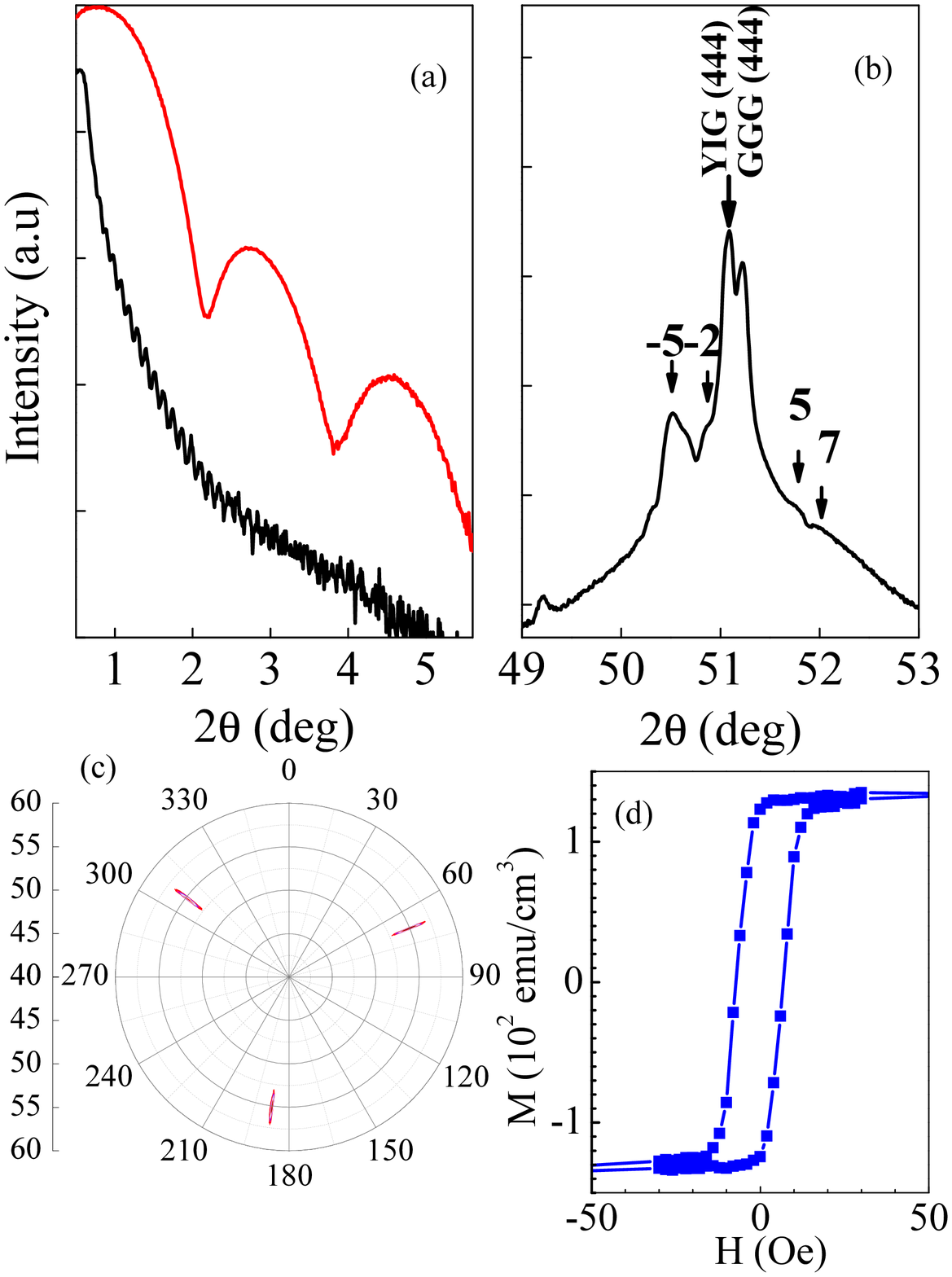}}
\caption{}
\label{Fig1}
\end{center}
\end{figure}

\begin{figure}[p]
\begin{center}
\resizebox*{15 cm}{!}{\includegraphics*{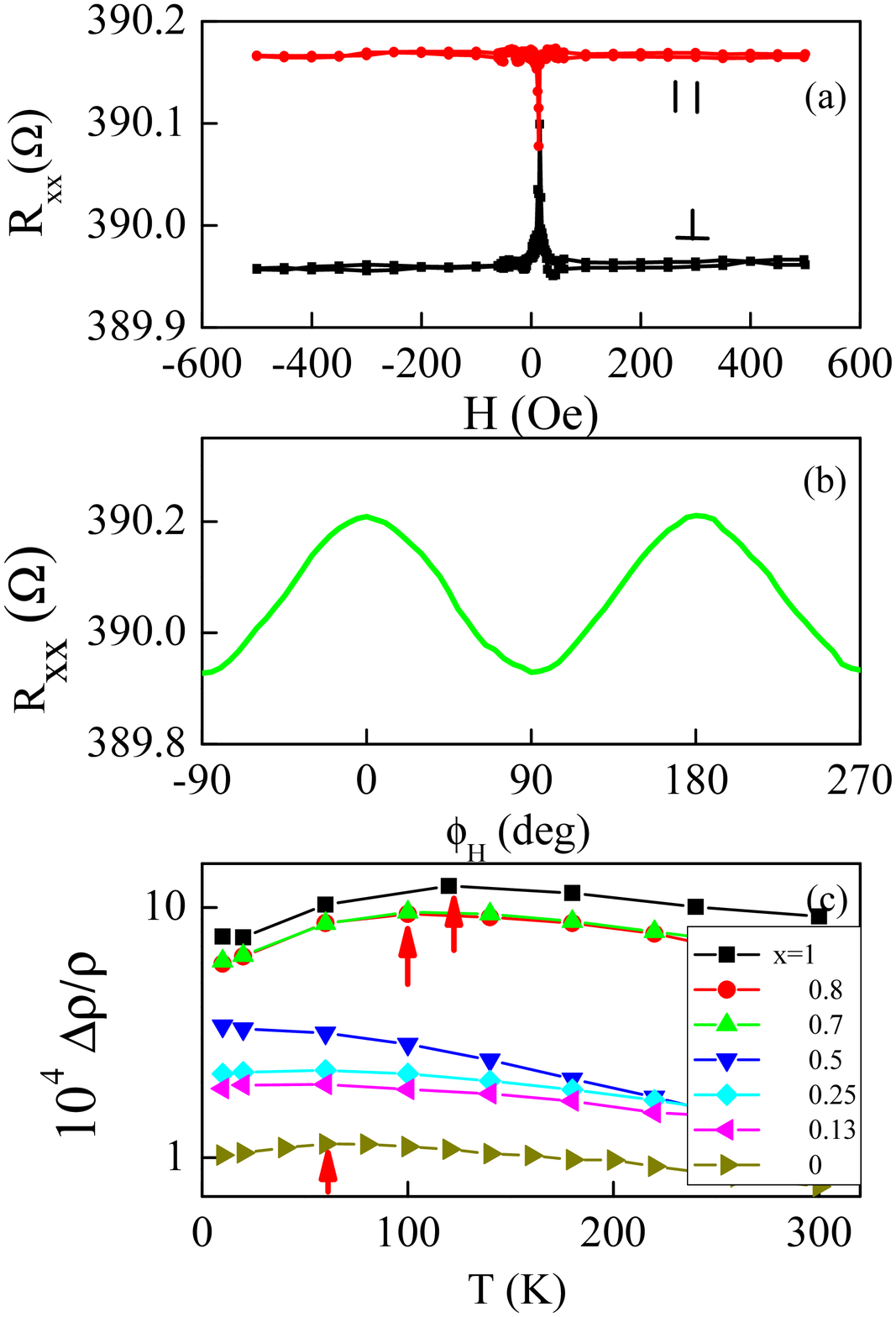}}
\caption{}
\label{Fig2}
\end{center}
\end{figure}

\begin{figure}[p]
\begin{center}
\resizebox*{15 cm}{!}{\includegraphics*{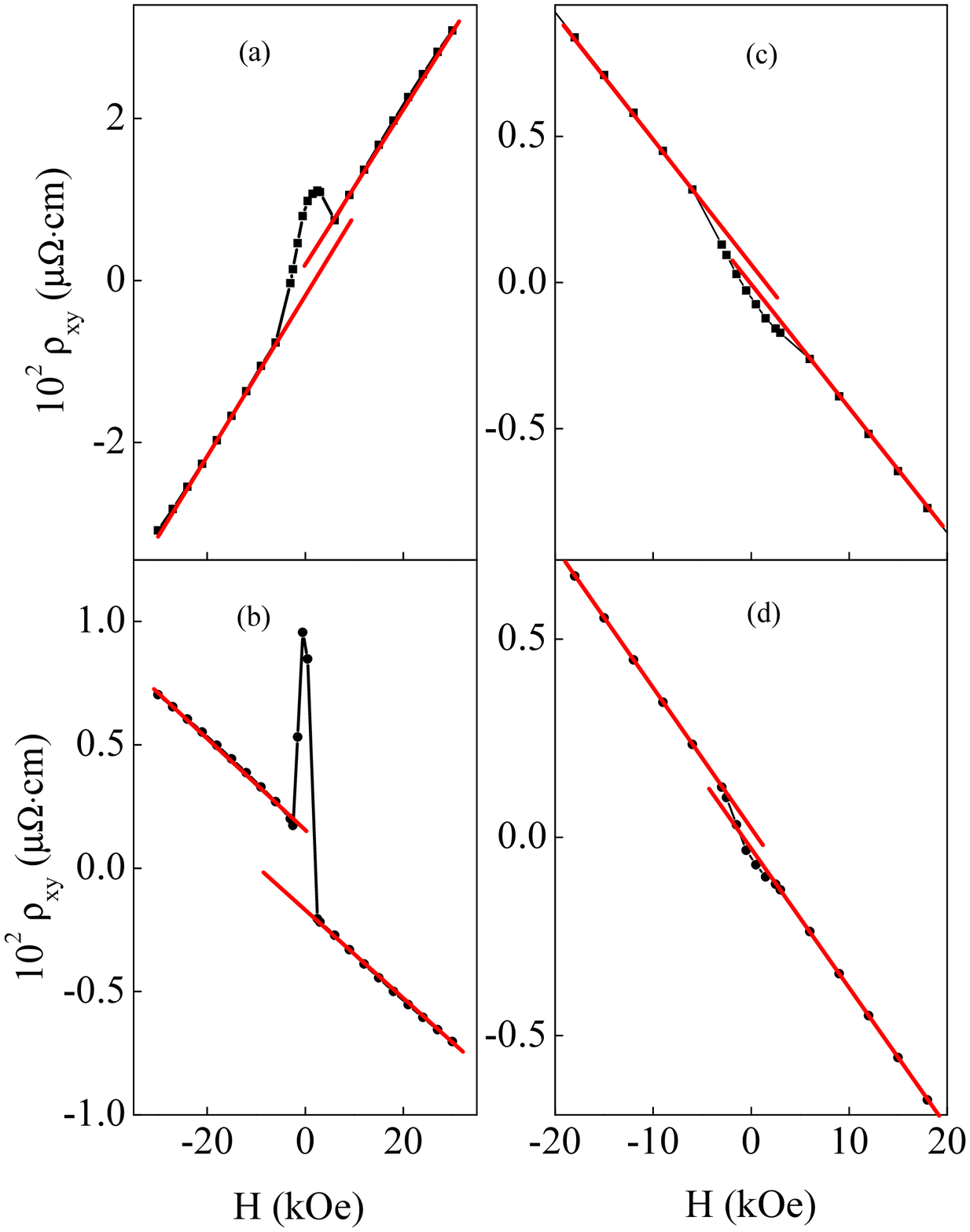}}
\caption{}
\label{Fig3}
\end{center}
\end{figure}

\begin{figure}[p]
\begin{center}
\resizebox*{15 cm}{!}{\includegraphics*{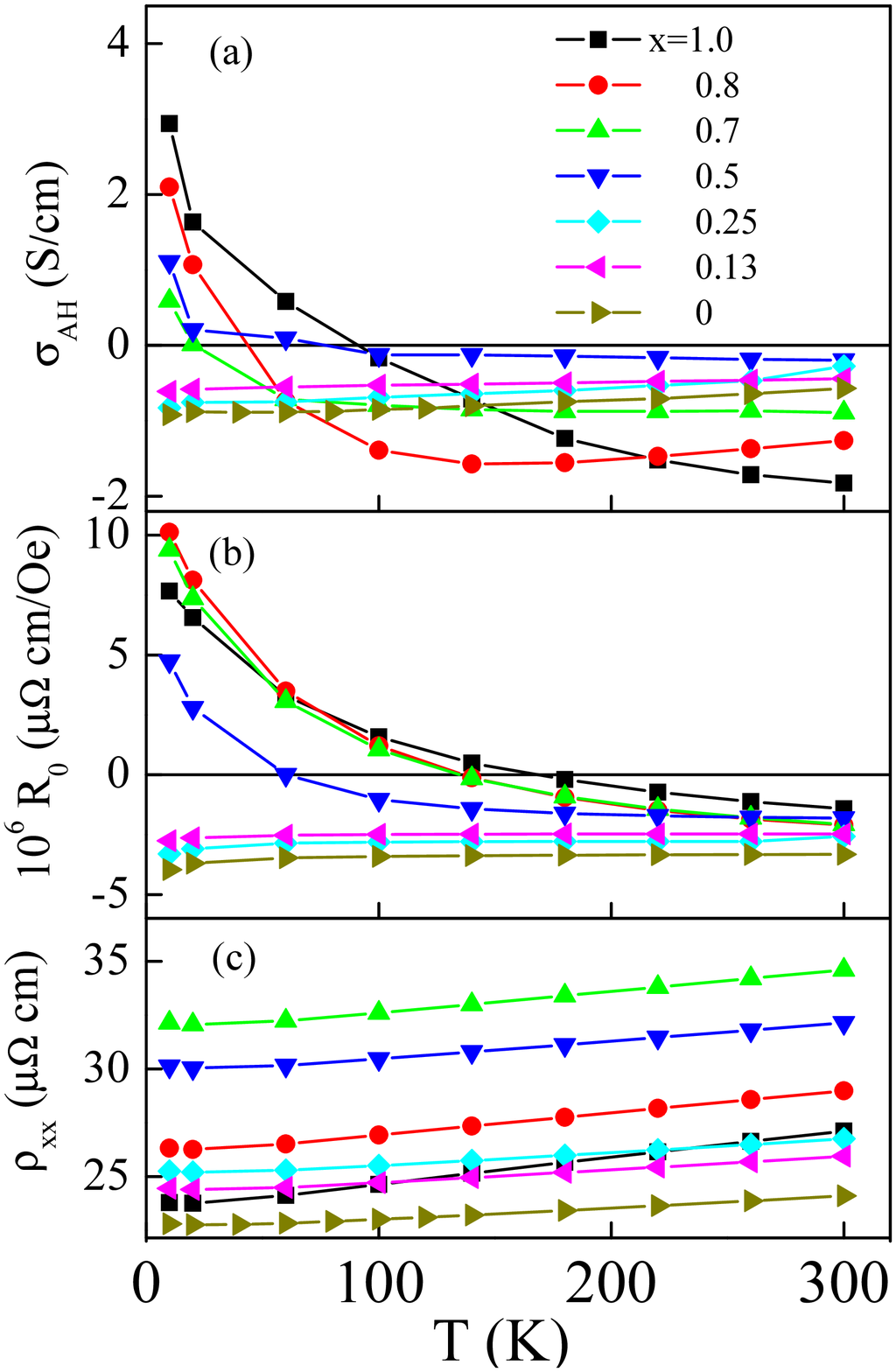}}
\caption{}
\label{Fig4}
\end{center}
\end{figure}

\begin{figure}[p]
\begin{center}
\resizebox*{15 cm}{!}{\includegraphics*{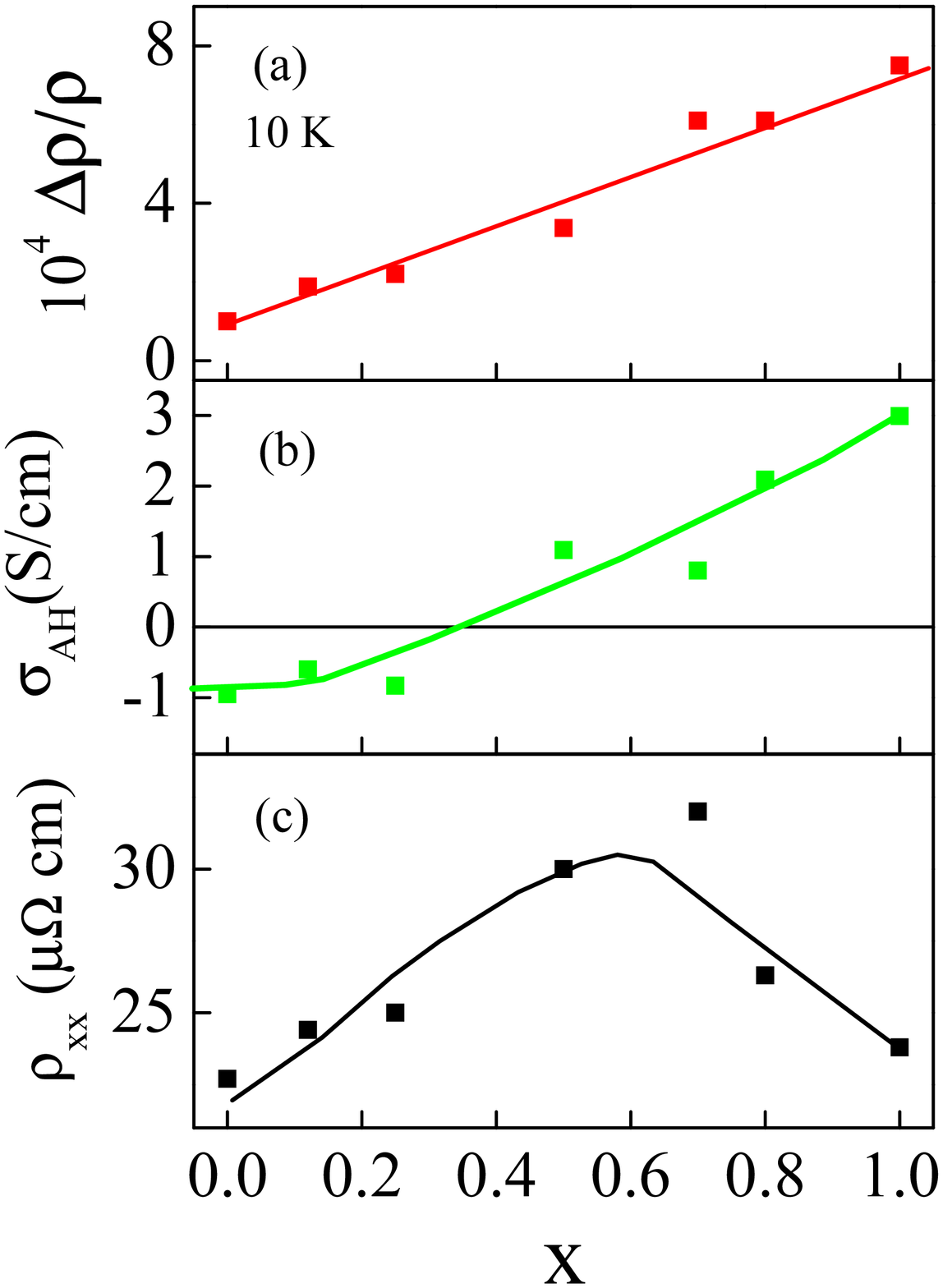}}
\caption{}
\label{Fig5}
\end{center}
\end{figure}

\end{document}